\documentclass[preprint,pra,aps,showpacs,preprintnumbers,amsmath,amssymb,eqsecnum]{revtex4}

\def\la{{\langle}}
\def\ra{{\rangle}}
\def\a{{\alpha}}
\def\b{{\beta}}

\def\e{{\epsilon}}
\def\l{{\lambda}}

\def\x{{\mathbf x}}

\def\p{\mathbf{p}}

\def\a{\alpha}
\def\b{\beta}

\def\au{{\underline \alpha}}

\def\half{\frac{1}{2}}
\def\pp{\prime\prime}

\newcommand\beq{\begin{equation}}
\newcommand\eeq{\end{equation}}
\newcommand\bea{\begin{eqnarray}}
\newcommand\eea {\end{eqnarray}}

\begin{document}


\title{Invariant Class Operators in the Decoherent \\ Histories
Analysis of Timeless Quantum Theories}

\author{J.J.Halliwell}%

\author{P.Wallden}

\affiliation{Blackett Laboratory \\ Imperial College \\ London SW7
2BZ \\ UK }

\date{\today}

\begin{abstract}
The decoherent histories approach to quantum theory is applied to
a class of reparametrization invariant models, which includes
systems described by the Klein-Gordon equation, and by a
minisuperspace Wheeler-DeWitt equation. A key step in this
approach is the construction of class operators characterizing the
questions of physical interest, such as the probability of the
system entering a given region of configuration space without
regard to time. In non-relativistic quantum mechanics these class
operators are given by time-ordered products of projection
operators. But in reparametrization invariant models, where there
is no time, the construction of the class operators is more
complicated, the main difficulty being to find operators which
commute with the Hamiltonian constraint (and so respect the
invariance of the theory).
Here, inspired by classical considerations, we put forward
a proposal for the construction of such class operators for a
class of reparametrization-invariant systems. They consist of
continuous infinite temporal products of Heisenberg picture
projection operators. We investigate the consequences
of this proposal in a number of simple models and also
compare with the evolving constants method.

\end{abstract}

\maketitle

\section{Introduction}

\subsection{Opening Remarks}

A problem attracting some interest in recent years concerns
the quantization of simple cosmological models which possess no
intrinsic time parameter, and which are described by an equation
of the Wheeler-DeWitt type
\beq
H \Psi = 0
\label{1.0}
\eeq
The absence of a time parameter together with the associated
reparametrization invariance represent a particular challenge to
conventional methods of quantization and interpretation and it has
proved surprisingly difficult to extract probabilities from
the wave function. Two particular approaches have made interesting
progress in this area: the evolving constants method \cite{Rov1,Rov2,Mar1,Haj,MRT,Mon,GaPo},
and the decoherent
histories approach
\cite{Har3,HaMa,HaTh1,HaTh2,CrHa,AnSa}.
The aim of this paper is to develop further the
decoherent histories quantization of these ``timeless''
theories described by an equation of the form Eq.(\ref{1.0}).

\subsection{The Decoherent Histories Approach}

We first briefly review the decoherent histories approach in non-relativistic
quantum theory described by a Schr\"odinger equation
\cite{GH1,GH2,Gri,Omn1,Omn2,Hal1,Hal2}.
In the decoherent histories approach probabilities are assigned to
histories via the formula,
\begin{equation}
p (\a_1, \a_2, \cdots ) = {\rm Tr} \left( C_{\au} \rho
C_{\au}^{\dag} \right)
\label{1.1}
\end{equation}
where $C_{\au} $ denotes a time-ordered string of projectors at
times $t_1 \cdots t_n$,
\begin{equation}
C_{\au} = P_{\a_n} (t_n) \cdots P_{\a_2} (t_2) P_{\a_1} (t_1)
\label{1.2}
\end{equation}
and $ \au $ denotes the string $\a_1, \a_2, \cdots \a_n $. The
projections operators are in the Heisenberg picture,
\begin{equation}
P_{\a_k} (t_k) = e^{ i H t_k} P_{\a_k} e^{ - i H t_k}
\label{1.3}
\end{equation}
where the projectors satisfy
\beq
\sum_{\a} P_{\a} = 1
\eeq
and
\beq
P_{\a} P_{\b} = \delta_{\a \b} P_{\a}
\eeq
We are
interested in sets of histories which satisfy the condition of
decoherence, which is that decoherence functional
\begin{equation}
D(\au,\au') = {\rm Tr} \left( C_{\au} \rho C_{\au'}^{\dag} \right)
\label{1.4}
\end{equation}
is zero when $\au \ne \au' $. Decoherence implies the weaker
condition that $ {\rm Re} D(\au,\au') = 0 $ for $\au \ne \au' $,
and this is equivalent to the requirement that the above
probabilities satisfy the probability sum rules. We normally
work with the stronger condition of decoherence, which is
related to the existence of records, corresponding to generalized
measurements \cite{GH2,Halrec}.

Now some simple observations relevant to what follows.
The class operators Eq.(\ref{1.2}) defined above satisfy
\begin{equation}
\sum_{\au} C_{\au} = 1
\label{1.5}
\end{equation}
However, in non-relativistic quantum mechanics, one can equally well
define the class operators to be
\begin{equation}
C_{\au}= P_{\a_n} e^{ - i H (t_n - t_{n-1}) } P_{\a_{n-1}} \cdots
e^{ - iH (t_2 - t_1) } P_{\a_1}
\label{1.6}
\end{equation}
as long as the initial density matrix is redefined to absorb a
unitary evolution factor (the unitary factors at the final time
cancel out in the decoherence functional). This alternative class
operator satisfies
\begin{equation}
\sum_{\au} C_{\au} = e^{ - i H (t_n - t_1) }
\label{1.7}
\end{equation}
The distinction between these two class operators is trivial in
non-relativistic quantum mechanics but not so in
reparametrization-invariant theories where one has to ask afresh
what a class operator actually is. The definition Eq.(\ref{1.6})
with the property Eq.(\ref{1.7}) views the class operator as the
decomposition of a propagator, and is best thought of in terms of
a restricted sum over paths in a path integral. The definition
Eq.(\ref{1.2}) with the property Eq.(\ref{1.5}), on the other
hand, views a class operator as the generalization of a projection
operator since clearly it would be a projection operator if all
the projections at different times commute. The difference between
these two views is irrelevant in non-relativistic quantum
mechanics but can have a significant influence when it comes to generalizations
to reparametrization-invariant theories, as we shall see shortly.

\subsection{Decoherent Histories for Systems without Time}

The structure of the decoherent histories approach is very general
and readily applies to a wide variety of situations, provided one
specifies a number of things in the construction of the decoherence
functional Eq.(\ref{1.4}),
such as the inner product structure
and the form of the class operators. Here we are concerned with
reparametrization-invarariant theories which are
characterized by a constraint equation of the form
\begin{equation}
H | \Psi \ra = 0
\label{1.8}
\end{equation}
where $H$ is usually quadratic in all the momenta.
Important examples are the Klein-Gordon equation of relativistic
quantum mechanics and the Wheeler-DeWitt equation of quantum cosmology.
The solutions
to this equation are usually not normalizable in the usual
Schr\"odinger inner product, so we use instead the so-called
induced (or Rieffel) inner product \cite{Rie}. This involves
first considering eigenstates of $H$,
\beq
H | \Psi_{Ek} \rangle = E | \Psi_{Ek} \rangle
\eeq
where $k$ is a degeneracy label. The spectrum of $H$ is typically
continuous and in the usual inner product we have
\beq
\langle \Psi_{Ek} | \Psi_{E'k'} \rangle = \delta (E - E') \delta (k-k')
\eeq
The induced inner product between the eigenstates with the same $E$
(including $E=0$)
is then defined, loosely speaking, by dropping the $\delta$ function
in $E$, that is
\beq
\langle \Psi_{Ek} | \Psi_{Ek'} \rangle_I = \delta (k-k')
\eeq
In practical terms, this means working with the usual inner product,
regularizing all expressions by working with eigenstates of $H$
and then dropping $\delta (E - E')$ at the end.

It is also useful to note that eigenstates of $H$ may be written
\beq
| \Psi_E \rangle = \delta ( H - E) | \phi \rangle
\label{1.10}
\eeq
for some fiducial state $ | \phi \rangle$. Then
\beq
\langle \Psi_E | \Psi_{E'} \rangle = \delta (E - E') \langle \phi | \delta ( H - E)
| \phi \rangle
\eeq
so the induced inner product between two eigenstates with the same eigenvalue is
\beq
\langle \Psi_E | \Psi_{E} \rangle_I =  \langle \phi | \delta ( H - E)
| \phi \rangle = \langle \Psi_E | \phi \rangle
\label{1.10b}
\eeq
See Refs.\cite{HaTh1,HaTh2,HaMa} for applications similar to those
considered here.

We are interested in the construction of the class operators for systems
of this type. The key property of reparametrization-invariant theories is that
they generally do not possess a variable to play the role of time, hence all
questions that one asks about the system must not refer to time in any way.
We will concentrate on the following useful question:
given that the system's state satisfies the constraint equation,
what is the probability of finding the system in a region $\Delta$ of
configuration space, without regard to time? The question is clearly
a sensible one classically,
since the system has a number of classical trajectories and one can
ask what proportion of them pass through the region in question.
Moreover, classically, it is also a reparametrization-invariant question,
since an entire classical trajectory is a reparametrization-invariant
object \cite{HaTh2}.
To answer this question in the quantum case we need to find a suitable
class operator.

It is generally held that for reparametrization-invariant theories,
the most significant class of physical questions involve operators
which commute with the Hamiltonian \cite{DeW,Mar1,Mar2,Rov1,Rov2,Rov3}.
These are referred to as
``observables'' and are the analogues of gauge-invariant quantities
in gauge theories. This issue is not without debate and subtlety
in the case of theories invariant under reparametrizations \cite{HalH},
but in this paper we will go along with this general idea.
We therefore seek a class operator $C_{\Delta}$ satisfying
\begin{equation}
[ H, C_{\Delta} ] = 0
\label{1.11}
\end{equation}
and which corresponds to the statement that the system passes through
the region $\Delta$ without reference to time.
With reference to the discussion of Section IB, we adopt the
projection operator view of the class operator, so the $C_{\Delta}$
becomes the identity when $\Delta$ becomes the entire configuration
space and is also a projector when everything commutes.

On the other hand, if we looked instead for a class operator which is
the analogue of the propagator form Eq.(\ref{1.6}), then $C_{\Delta}$ become
$\delta (H)$ when $\Delta$ becomes the entire configuration
space (although this is essentially the identity
when operating on solutions to the constraint equation). We then
expect that class operator to satisfy the constraint equation
\begin{equation}
H C_{\Delta} = 0
\label{1.12}
\end{equation}
However, this propagator
viewpoint naturally leads to a path integral construction
which, in earlier works, was found difficult to reconcile with the constraint
equation, Eq.(\ref{1.12}) \cite{HaTh1,HaTh2}. In this paper we will
therefore concentrate on the projection operator form, the generalization
of Eq.(\ref{1.2}), and reparametrization invariance is easily maintained.

On a more practical note, in order to evaluate the decoherence functional and
probabilities, we will need to evaluate expressions of the form
$ \langle \Psi | A | \Psi \rangle$
where $A$ commutes with $H$ and $ | \Psi \rangle$
is an eigenstate of $H$. Using Eq.(\ref{1.10}), we have
\bea
\langle \Psi_E | A | \Psi_{E'} \rangle &=&
\langle \phi | \delta (H - E) A \delta ( H - E') | \phi \rangle
\nonumber \\
&=& \delta (E - E') \langle \phi | \delta (H - E) A | \phi \rangle
\eea
so in the induced inner product we have
\bea
\langle \Psi_E | A | \Psi_{E} \rangle_I
&=& \langle \phi | \delta (H - E) A | \phi \rangle
\nonumber \\
&=& \langle \Psi_E | A | \phi \rangle
\label{1.12c}
\eea
For example, using this formula, the decoherence functional
(with class operators commuting with $H$) is conveniently written
\bea
D(\a, \a') &=& \langle \Psi_E | C_{\a'}^{\dag} C_{\a} | \Psi_E \rangle_I
\nonumber \\ &=&
\langle \phi | C_{\a'}^{\dag} \delta ( H - E) C_{\a} | \phi \rangle
\eea

\subsection{This Paper}

In Section 2 we describe the earlier attempts to construct class operators
for timeless models and the associated mathematical machinery that we will need
here.
In Section 3 we describe our proposal
for new class operators which are compatible with the constraint.
They consist of infinite products in time of projection
operators in the Heisenberg picture.
We compare the decoherent histories approach with the ``evolving constants''
method in Section 4. In Section 5, we show that the decoherent histories approach
together with the new class operators gives sensible and expected results for the
non-relativistic particle in parametrized form. In Section 6 we compute the
class operators and decoherence functional for a simple one-dimensional example
and we apply this understanding to the relativistic particle in Section 7.
We look at some simple examples in two dimensions in Section 8. The very different
and simpler case of systems of harmonic oscillators is covered in Section 9.
We summarize and conclude in Section 10.

\section{Background}

We begin by describing the propagator viewpoint for the construction
of class operators for timeless systems. As stated, this has difficulties
in relation to the constraint,
but the details of the construction are important.
Recall that we are interested in the question, given that the system is
in an energy eigenstate, what is the
probability of finding the particle in a region $\Delta$ of
configuration space, without regard to time?

We will consider a system whose $d$-dimensional configuration
space is $\mathbb{R}^d$ and it will generally be useful to denote their
coordinates by a vector $\x$, although when talking about the
relativistic particle, we will use the usual notations $x$ or
$x^{\mu}$. The propagator approach to defining the
class operators is to define them by summing over all paths in the
configuration space between given end points which pass through
the region $\Delta$ \cite{HaTh1,HaTh2,HaMa,Har3}.
In this approach, the class operator of interest is therefore
given by
\begin{equation}
C_{\Delta}(\x^{\pp},\x') = \int_{-\infty}^{\infty} d T \
g_{\Delta} (\x^{\pp}, T | \x', 0 )
\label{B1}
\end{equation}
The integrand is given by a standard path integral (of
non-relativistic type)
\begin{equation}
g_{\Delta} (\x^{\pp}, T | \x', 0 ) = \int {\cal D} \x \exp \left( i S
[\x(t)] \right)
\end{equation}
where the sum is over all paths from $\x'$ to $\x^{\pp}$ in time
$T$  which pass through $\Delta$ and $S[\x(t)]$ is an action of
the usual form
\begin{equation}
S[\x (t)] = \int_0^T dt \left( f_{ij} \dot x_i \dot x_j - V (\x ) \right)
\end{equation}
for some metric on the configuration space $f_{ij}$ (whose explicit
form will be unimportant in this section).
This definition seems reasonable since it is an obvious generalization
of Eqs.(\ref{1.6}), (\ref{1.7}). Also,
if we let the region $\Delta$
become the whole configuration space, then $g$ is a solution to
the Schr\"odinger equation, and, since the integration range of
$T$ is infinite, $C(\x^{\pp},\x')$ is a solution to the constraint
equation,
\begin{equation}
H C = 0
\end{equation}
(If we are interested in an eigenstate of energy $E$, then we may
assume that $E$ has been absorbed into the potential $V$).
However, there is a fundamental problem with this construction,
which is that the class operator does not appear to satisfy the
constraint equation everywhere, except for the case when $\Delta $ is the
whole configuration space.

To see this, it is necessary to go into more detail about the
construction of the above class operators. These details will also
be important for the construction of projector-type class operators
which commute with the constraint. We first introduce the
(provisional) class operator for not entering the region $\Delta$,
which is given by a restricted sum over paths that do not enter
$\Delta$,
\begin{equation}
C_r (\x^{\pp},\x') = \int_{-\infty}^{\infty} d T  \
g_r (\x^{\pp}, T | \x', 0 )
\label{res2}
\end{equation}
Here $ g_r (\x^{\pp}, T | \x', 0 ) $ is the non-relativistic
restricted propagator, defined by a sum over paths in fixed time
$T$ that do not enter $\Delta$. It vanishes when either end-point
is in $\Delta$ or on its boundary. Since the set of all paths between
the fixed end-points either pass through $\Delta$ or not, we have
\begin{equation}
g (\x^{\pp}, T | \x', 0 ) =  g_r (\x^{\pp}, T | \x', 0 ) + g_{\Delta} (\x^{\pp}, T | \x', 0 )
\end{equation}
and correspondingly
\begin{equation}
C (\x^{\pp},\x') = C_r (\x^{\pp},\x') + C_{\Delta} (\x^{\pp},\x')
\label{cr}
\end{equation}
Here, $C (\x^{\pp}, \x') $ denotes the sum over all paths, and in fact
\begin{equation}
C (\x^{\pp}, \x' ) = \la \x^{\pp} | \delta ( H) | \x' \ra
\end{equation}

There is a way of writing the restricted propagator which will be useful for later sections.
We introduce the projection operator $P$ onto the region $\Delta$,
\begin{equation}
P = \int_{\Delta} d^d x | \x \ra \la \x |
\end{equation}
together with the complementary projector $\bar P = 1 - P $ onto the region  $\bar \Delta $
outside
$\Delta$. Suppose we divide the time interval $[t',t^{\pp}]$ into discrete points,
$ t' = t_0 < t_1 < t_2 < \cdots t_{n-1} < t_n = t^{\pp} $, where $t_{k+1} - t_k = \delta t$.
We introduce the operator version $g_r (t^{\pp},t') $ of the restricted propagator,
$g_r (\x^{\pp}, t^{\pp} | \x', t' )$, so
\begin{equation}
g_r (\x^{\pp}, t^{\pp} | \x', t' ) = \la \x^{\pp} |  g_r (t^{\pp},t') | \x' \ra
\end{equation}
The operator version is then given by
\begin{equation}
g_r (t^{\pp}, t' ) = \lim_{\delta t \rightarrow 0 }  \ \bar P e^{ - i H (t_n-t_{n-1})}
\bar P e^{ - i H (t_{n-1}-t_{n-2})} \cdots
\bar P e^{ - i H (t_1-t_{0})} \bar P
\label{res}
\end{equation}
where the limit is $\delta t \rightarrow 0 $, $n \rightarrow \infty$ with
$ n \delta  t = (t^{\pp} - t') $ held constant. From this one can clearly see that
$ g_r (\x^{\pp}, t^{\pp} | \x', t' ) $ vanishes if either end-point is in $\Delta$.
One may also see that it does not quite satisfy the Schr\"odinger equation, but
satisfies instead,
\begin{equation}
\left( i \frac { \partial} {\partial t^{\pp}} - H \right) g_r ( t^{\pp}, t' )
= [ \bar P, H ] g_r (t^{\pp},t')
\end{equation}
Because $\bar P$ is a projection operator, in the $\x$ representation the right-hand side
consists only of $\delta$-functions on the boundary of $\Delta$. So the restricted propagator
almost satisfies the Schr\"odinger equation, but just fails at the boundary. Correspondingly,
when used to construct the class operator $C_r (\x^{\pp},\x') $ in Eq.(\ref{res2}), it fails to satisfy
the constraint because of $\delta$-functions on the boundary. Consequently, $C_{\Delta}$
also fails to satisfy the constraint, because of Eq.(\ref{cr}).

It should be noted that constructions such as Eq.(\ref{res2}) can, in fact, be argued to be
reparametrization invariant and one might therefore expect that it satisfies
the constraint equation. The fact that is does not quite satisfy the constraint is related
to subtle differences between the way reparametrizations act in configuration space versus
phase space \cite{HalH}.

Another useful formula for the construction of these class operators is the
so-called path decomposition expansion \cite{PDX,Hal3,HaOr}.
The propagator $ g_{\Delta} (\x^{\pp}, t^{\pp}| \x', t' ) $
is given by a sum over paths which enter the region $\Delta$. These paths may be partitioned
according to the time $t_c$ and place $\x_c$ at which they cross the boundary $\Sigma$ of
$\Delta$ for the first time. The crossing propagator may then be written,
\begin{equation}
g_{\Delta} ( \x^{\pp}, t^{\pp} | \x', t' )
= \int_{t'}^{t^{\pp}} dt_c \int_{\Sigma} d^{d-1} \x_c
\ g( \x^{\pp}, t^{\pp} | \x_c, t_c )
\ {i  \over 2 m} {\bf n} \cdot {\bf \nabla}
g_r ( \x_c, t_c | \x', t' )
\label{pdx}
\end{equation}
where the normal ${\bf n}$ points towards the restricted propagation region.
Although the restricted propagator $g_r$ vanishes on $\Sigma$, its normal derivative
does not (if defined by first taking the derivative and then letting
$\x_c$ approach the surface from within the restricted propagation region).
In fact the combination $ ({i  / 2 m}) {\bf n} \cdot {\bf \nabla}
g_r ( \x_c, t_c | \x', t' ) $ represents a sum over paths which do not cross $\Sigma$
but end on it.

The path decomposition expansion was used in Ref.\cite{HaTh1} to compute class
operators corresponding to crossings of a spacelike surface in relativistic
quantum mechanics. As stated above, the class operators constructed using the
above methods failed to satisfy the constraint.
However, following a suggestion in Ref.\cite{HaMa},
it was shown that operators satisfying the constraint and yielding sensible results could
be obtained by some simple and physically reasonable modifications.
But this procedure is rather ad hoc and it
was not clear how to turn it into a general definition of the class operator.

Another clue as to how class operators should be constructed
was found in Ref.\cite{HaTh2}, which considered general
minisuperspace models and attempted to construct class operators
for them. The starting point was the construction of probabilities
for timeless coarse grainings in the classical theory. Suppose we
have a classical theory described by a phase space probability
distribution function $w (\p, \x)$ satisfying
\beq
\{ H, w \} = 0
\eeq
(the classical analogue of the constraint equation).
Let $f_{\Delta} (\x)$ denote
the characteristic function of the region $\Delta$, so is $1$ or
$0$, depending on whether $\x$ is inside or outside $\Delta$. Now
introduce the classical solution $ \x^{cl} (t)$ passing through
the phase space point $\p, \x$. Then the quantity
\begin{equation}
\tau_{\Delta} = \int_{-\infty}^{\infty} dt \ f_{\Delta} ( \x^{cl}
(t) )
\label{Y.2}
\end{equation}
is the amount of parameter time spent by the trajectory in the
region $\Delta$. This quantity has the important property
that it has vanishing Poisson bracket with the Hamiltonian,
\begin{equation}
\{ H, \tau_{\Delta} \} = 0
\end{equation}
so is
a classical observable. To determine whether or not the trajectory passes
through $\Delta$, we only need to know if $ \tau_{\Delta}$ is
positive or zero. It follows that the probability of entering
$\Delta$ is given by
\begin{equation}
p_{\Delta} = \int d^d p d^d x \ w (\p, \x ) \ \theta \left(
\int_{-\infty}^{\infty} dt \ f_{\Delta} (\x^{cl} (t) ) - \e
\right)
\label{Y.3}
\end{equation}
Here, $\epsilon$ is a small parameter which goes to zero through
positive values, and is included to avoid ambiguities in the
$\theta$-function at zero argument. The whole expression is
invariant under reparametrizations, since each part of it is.
Similarly, the probability for not entering
the region is obtained by flipping the sign of the argument in the $\theta$-function.
(Note that there is an issue of normalization of $w(\p, \x)$ in Eq.(\ref{Y.3}), since
$w(\p, \x)$ is constant along the classical trajectories. This issue is in fact
resolved by the normalization in the analogous quantum case, as discussed in
Ref.\cite{HaTh2}).

Inspired by the classical case, it was suggested in Ref.\cite{HaTh2} that in the
quantum case, the class operator in the semiclassical
approximation is given by
\begin{equation}
C_{\Delta} (\x_f, \x_0) = \theta \left( \int_{-\infty}^{\infty} dt
\ f_{\Delta} (\x_0^f (t) ) -\e \right) \ B(\x_f, \x_0) \ e^{ i
A(\x_f, \x_0)}
\label{Y.4}
\end{equation}
where $B e^{iA}$ is the usual unrestricted semiclassical
propagator, and $\x_0^f (t)$ denotes the classical path connecting
$\x_0$ to $\x_f$. This object satisfies the constraint in the
semiclassical approximation and gave sensible results,
but no fully quantum version was given.

Note that the classical and semiclassical results Eqs.(\ref{Y.3}), (\ref{Y.4})
involve entire classical trajectories, not trajectories of finite length between
fixed end-points as indicated by constructions such as Eq.(\ref{B1}). This
is significant since, as argued previously, a whole classical trajectory
is reparametrization invariant, whereas a section of classical
trajectory is not \cite{HaTh2}. Hence one of the key ideas in the quantum theory is to get
away from propagation between fixed end-points and towards objects which
capture the idea of an entire trajectory, as we will see shortly.


\section{New Class Operators}

Given the above background and difficulties, the question now is how to
define class operators that commute with the constraint and that give sensible
semiclassical results. In this section we will focus on the case in which the unphysical
parameter time $t$ takes an infinite range. The special case in which the parameter
time is periodic (bound systems) will be treated in Section 9.

A useful hint towards constructing class operators
comes from the $\theta$-function used in the expressions
(\ref{Y.3}) and (\ref{Y.4}). Suppose we are interested in the probability of not entering $\Delta$.
Let is write the appropriate $\theta$-function in terms of its Fourier transform,
\begin{equation}
\theta \left( \e  - \tau_{\Delta} \right) = \int \frac{ dk} {ik} e^{ i k ( \e - \tau_{\Delta}) }
\label{3.1}
\end{equation}
where $\tau_{\Delta}$ is given by Eq.(\ref{Y.2}). As stated, this object is
reparametrization-invariant in that
it has vanishing Poisson bracket with $H$.
Now consider a discretized version of the time
integral for $\tau_{\Delta}$, so we split the time into small intervals of size
$\delta t $, and we have
\begin{equation}
\tau_{\Delta} \ \approx \  \delta t \sum_{n=-\infty}^{\infty} f_{\Delta} ( \x (t_n) )
\label{3.2}
\end{equation}
with exact agreement with the original expression in the limit $\delta t \rightarrow 0 $.
This then means that the $\theta$-function is given by the continuum limit of
the expression
\begin{equation}
\theta \left( \e - \tau_{\Delta} \right)
= \int \frac{dk} {ik} \ e^{ i k \e } \ \prod_{n=-\infty}^{\infty}
\exp \left( { - i k \delta t f_{\Delta} (t_n) } \right)
\label{3.3}
\end{equation}
But now $f_{\Delta}$ is a characteristic function, so is $0$ or $1$. It follows
that
\begin{equation}
\exp \left( { - i k \delta t f_{\Delta} } \right)
= f_{\bar \Delta} + e^{ - i k \delta t } f_{\Delta}
\label{3.4}
\end{equation}
where $ f_{\bar \Delta} = 1 - f_{\Delta}$ is the characteristic
function for the region $\bar \Delta$ outside $\Delta$, and therefore
\begin{equation}
\theta \left( \e - \tau_{\Delta} \right)
= \int \frac{dk} {ik} \ e^{ i k \e } \ \prod_{n=-\infty}^{\infty}
\left[ f_{\bar \Delta} (t_n) +  e^{ - i k \delta t } f_{\Delta} (t_n) \right]
\label{3.5}
\end{equation}
This result has a very appealing form. When the product is expanded out, we get sums of
products of the characteristic functions $ f_{\Delta}$ and $f_{\bar \Delta}$, so the
first term, for example, is the continuum limit of
\begin{equation}
\prod_{n=-\infty}^{\infty} f_{\bar \Delta} (t_n)
\label{3.6}
\end{equation}
This quantity is clearly equal to $1$ for
a classical history in which the particle is outside $\Delta$
at every point along its trajectory and is zero otherwise.
The other terms involve similar histories including the function
$f_{\Delta}$, so these are histories which enter $\Delta$ for some of the time.
The integration over $k$ produces a $\theta$-function ensuring that only histories
which spend time less than $\e$ in the region $\Delta$ are included. In particular,
as $\e \rightarrow 0$, the only term that is left is the first term, Eq.(\ref{3.6}).
This is reparametrization-invariant because the expression it was derived from is.
The important conclusion from this is that we might therefore expect to obtain
reparametrization-invariant class operators in the quantum theory by taking infinite
products of projection operators.

Turning now to the quantum theory, it is well-known that operators commuting with $H$
can be constructed using the formula,
\begin{equation}
A = \int_{-\infty}^{\infty} dt \ B(t)
\label{3.7}
\end{equation}
where
\begin{equation}
B (t) = e^{i H t } B e^{ - i H t}
\label{3.8}
\end{equation}
The operator $A$ commutes with $H$ because
\begin{equation}
e^{ i H \tau} A e^{ - i H \tau} = \int_{-\infty}^{\infty} dt \ B(t + \tau)
= \int_{-\infty}^{\infty} dt \ B(t) = A
\label{3.9}
\end{equation}
so
\begin{equation}
[H,A] = 0
\label{3.10}
\end{equation}
Suppose we let $B = \ln b $. Then, very loosely speaking
\begin{equation}
A = \int_{-\infty}^{\infty} dt \ B(t)  = \ln \left( \prod_{t=-\infty}^{\infty} b(t) \right)
\label{3.11}
\end{equation}
That is, to the extent that the continuous product over $t$ is defined, we expect
that operators of the form
\begin{equation}
\prod_{t=-\infty}^{\infty} b(t)
\label{3.12}
\end{equation}
will commute with $H$.

Given these motivational remarks, we now give the new proposal for class
operators for trajectories that never enter the region $\Delta$. As before, denote
by $P$ the projector onto $\Delta$ and $\bar P$ the projector onto
the outside of $\Delta$. Then our proposal for the class operator
for trajectories not entering $\Delta$ is the time-ordered infinite product,
\begin{equation}
C_{\bar \Delta} = \prod_{t=-\infty}^{\infty} \bar P (t)
\label{3.13}
\end{equation}
To define this more precisely, we first consider the product of
projectors at a discrete set of times, $ t' = t_0 < t_1 < t_2 < \cdots t_{n-1} < t_n = t^{\pp} $,
where $ t_{k+1} - t_k = \delta t $.
We define the intermediate quantity, $ C_{\bar \Delta} (t^{\pp},t')$ as the continuum limit
of the product of projectors,
\begin{equation}
C_{\bar \Delta} (t^{\pp},t') = \lim_{\delta t \rightarrow 0} \bar P (t_n) \dots \bar P(t_1) \bar P (t_0)
\label{3.15}
\end{equation}
where the limit is $n \rightarrow \infty$, $\delta t \rightarrow 0 $ with
$ t^{\pp} - t'$ fixed. Finally, the desired class operator is
\begin{equation}
C_{\bar \Delta} = \lim_{t^{\pp} \rightarrow \infty, t' \rightarrow -\infty} \ C_{\bar \Delta} (t^{\pp},t')
\label{3.16}
\end{equation}

This new class operator is clearly closely related to the restricted propagator defined
above, Eq.(\ref{res}) but differs by the presence of unitary evolution operators at either end. In particular, we have
\begin{equation}
C_{\bar \Delta} (t^{\pp},t') = e^{ i H t^{\pp}} \ g_r (t^{\pp},t') \ e^{ - i H t'}
\label{3.17}
\end{equation}
and therefore
\begin{equation}
C_{\bar \Delta} = \lim_{t^{\pp} \rightarrow \infty, t' \rightarrow -\infty} \  e^{ i H t^{\pp}} \ g_r (t^{\pp},t') \ e^{ - i H t'}
\label{3.18}
\end{equation}
which is the most useful form of the class operator. This is the main
result of this section.

As required, the new class operator commutes with $H$. This is implied by the construction,
but more explicitly, we have from Eq.(\ref{3.17})
\begin{equation}
e^{ i H s} C_{\bar \Delta} (t^{\pp},t') e^{ - i H s} = e^{ i H (t^{\pp}+s)}
\ g_r (t^{\pp},t') \ e^{ - i H (t'+s)}
\label{3.18b}
\end{equation}
This becomes independent of $s$ as $ t^{\pp} \rightarrow \infty$, $t' \rightarrow - \infty$,
hence
\begin{equation}
[ H, C_{\bar \Delta} ] = 0
\end{equation}

Note that there is no reason at this stage why one should not use a different
operator ordering of the projectors in Eq.(\ref{3.15}). (This issue will become
significant in the bound case treated later). Here, we investigate the
consequences of the chosen ordering, which appears to be the simplest, but
keeping in mind that a different choice may be appropriate.

The class operator Eq.(\ref{3.18})
is quite different from the original proposal for this class operator,
Eq.(\ref{res2}), in that it does not involve an integral over parameter time.
Furthermore, unlike Eq.(\ref{res2}), the new class operator $C_{\bar \Delta} (\x^{\pp},\x')$
defined in this way does {\it not} in general vanish when either end point is in
$\Delta$, so it is not perfectly localized in $\bar \Delta$.
In some sense, it corresponds to paths which do
not enter the region $\Delta$ but are allowed to enter it at infinite parameter time.
On the other hand, the new class operator is thoroughly compatible with the constraint
equation, since it commutes with $H$, whereas Eq.(\ref{res2}) does not quite
satisfy the constraint.

Generally, in the quantum theory, because the position operator does not commute
with $H$, there is an incompatibility between localization in configuration space
and the constraint equation. It is therefore necessary to make a choice as to
which of these two requirements should be given precedence.
The original proposal Eq.(\ref{res2}) has exact
spatial localization, but is not fully compatible with the constraint. The new
class operators are fully compatible with the constraint but are not perfectly
localized in configuration space. Hence the current approach gives precedence to
the constraint equation over localization.

Also, as noted earlier, the symmetry of reparametrization invariance is quite subtle
in that Eq.(\ref{res2}) can be argued to be invariant under the configuration space form
of reparametrizations, even though it is not fully compatible with the constraint.
The symmetry generated by $H$ is slightly larger than the configuration space form
of reparametrizations, so in the new class operators we are demanding a slightly more restrictive
notion of invariance than in Eq.(\ref{res2}).
It would be of interest to explore these subtle differences in greater detail.

The class operator $C_{\Delta}$ for entering the region $\Delta$ is now simply defined by
\begin{equation}
C_{\Delta} = 1 - C_{\bar \Delta}
\label{3.19}
\end{equation}
A more enlightening formula for it may however be obtained using the path decomposition
expansion, Eq.(\ref{pdx}). In particular, we clearly have
\begin{equation}
C_{\Delta} = \lim_{t^{\pp} \rightarrow \infty, t' \rightarrow -\infty} \  e^{ i H t^{\pp}} \ g_\Delta (t^{\pp},t') \ e^{ - i H t'}
\label{3.20}
\end{equation}
where $g_{\Delta} (t^{\pp},t')$ is defined by
\begin{equation}
g_{\Delta} (\x^{\pp}, t^{\pp} | \x', t' ) = \la \x^{\pp} |  g_{\Delta} (t^{\pp},t') | \x' \ra
\label{3.21}
\end{equation}
and the left-hand side is given by the path decomposition expansion, Eq.(\ref{pdx}).

It is not immediately clear from the definition of these class operators that they will
exist in all situations of interest. In particular, one would expect that the continuous
products over time and infinite limits will require careful attention. The proper mathematical framework
for handling these quantities is the continuous tensor product structure defined by
Isham et al. \cite{Ish}.
Here, we will proceed in a more informal way, and we will see by explicit computation in
specific examples that the class operator exists and gives reasonable results.
A more rigorous approach to quantizing models of this type is being pursued by Anastopoulos
and Savvidou \cite{AnSa}, using the structures developed by Isham et al. \cite{Ish}.
Future papers will address the connection between the present approach and these more rigorous
approaches.

\section{Comparison with the Evolving Constants Method}

The decoherent histories approach considered here for timeless theories
bears comparison with the evolving constants method of Rovelli
\cite{Rov1} (for further developments see Refs.\cite{Mar1,Kuc,Haj,Har10,MRT,Mon,GaPo}.
In that method, one
constructs operators commuting with the constraint corresponding to physically
interesting questions. For these operators, one may construct projections $P_{\a}$
onto ranges of the spectrum and the probabilities then have the
usual form ${\rm Tr} ( P_{\a} \rho)$.

For example, suppose the system is a free particle in two
dimensions, with Hamiltonian
\beq
H = \frac {p_1^2} {2m} + \frac {p_2^2} {2m}
\eeq
Suppose we are interested in the question, what is the value of $x_1$ when
$x_2 = \tau$? The corresponding evolving constants variable is
\beq
X_1 (\tau)  = \int_{-\infty}^{\infty} ds \ x_1 (s)\ \frac {dx_2 (s)} {ds} \ \delta
( x_2 (s) - \tau)
\eeq
where
\beq
x_i (s) = x_i + \frac {p_i} {m} s
\eeq
for $i=1,2$. This clearly has vanishing Poisson bracket with $H$. Classically, the integral
over $s$ may be carried out with the result
\beq
X_1 (\tau) =  x_1 + \frac {p_1} {p_2} ( \tau - x_2 )
\eeq
The $ 1/p_2 $ factor presents difficulties in turning this into a self-adjoint
operator, and as a consequence the spectrum of states is not orthonormal.
One can still construct a POVM onto a range of the spectrum
but it will not be an exact projector, since it will not satisfy
$P_{\a} P_{\b} = 0 $ for $\a \ne \b $. This leads to a kind of imprecision
in their definition. These POVMs however, may still be useful,
in the same way the phase space localized quasi-projectors are useful.

To compare with the decoherent histories approach, suppose we are interested
in the probability of entering or not entering a region $\Delta$.
Consider therefore the expression Eq.(\ref{Y.2}) for the parameter time
spent in $\Delta$, which in this simple model is
\beq
\tau_{\Delta} = \int_{-\infty}^\infty ds \  f_{\Delta} (x_1 (s), x_2 (s))
\label{EC5}
\eeq
Eq.(\ref{EC5}) may be written
\beq
\tau_{\Delta} =  \int_{-\infty}^\infty ds \int_{\Delta} dy_1 dy_2
\ \delta (x_1 (s) - y_1) \ \delta ( x_2 (s) - y_2)
\eeq
The $s$ integral may then be done with the result
\beq
\tau_{\Delta} = \int_{\Delta} dy_1 dy_2 \ \frac { m } {p_2}
\ \delta ( x_1 + \frac {p_1} {p_2} ( y_2 - x_2) - y_1 )
\eeq
Importantly, the result depends only on the evolving constants
operator $X_1 ( y_2) $ and on $p_2$ (both of which commute with $H$).
Hence in the evolving constraints approach one would consider the spectrum
of the operator $\tau_{\Delta}$, using what is known about the spectrum
of $X_1 (y_2)$ and $p_2$, and attempt to construct a projector or POVM
onto ranges of the spectrum of $\tau_{\Delta}$ (bearing in mind the difficulties
noted above with self-adjointness). We can then find the probability of
not entering $\Delta$ using the quasi-projector $\theta ( \epsilon - \tau_{\Delta })$.

In the decoherent histories approach, we also take Eq.(\ref{EC5}) as
the starting point. However, from that we deduce
the classical expression Eq.(\ref{3.6}), which may be used
for computation of the probability
of not entering $\Delta$. This is the starting point for the quantum theory,
and in particular, it
inspires the construction of class operators in terms of products of projection
operators, as described in the previous
section. Importantly, class operators are {\it not} required to be self-adjoint
operators, which in some sense means there is more freedom in the decoherent
histories approach.
On the other hand, in the decoherent histories approach probabilities cannot
be defined in general, unless there is decoherence, so in this sense the
theory is more restrictive than the evolving constants method.

The two resulting quantum theories are clearly quite different. However,
what they have in common is that they take starting points which are classically equivalent.
A more detailed comparison of these two approaches will be undertaken elsewhere.

\section{The Parametrized Non-Relativistic Particle}

When developing a quantization scheme for parametrized systems, one of the most important simple
systems to apply it to is the parametrized non-relativistic particle. This is because its quantum
theory is standard non-relativistic quantum mechanics and it is therefore easy to check whether
the expected results are reproduced by the methods described in Section 3.

\subsection{The Parametrized Particle and its Quantization}

The parametrized particle is the usual non-relativistic particle but with the time
coordinate $t$ raised to the status of a dynamical variable,
with conjugate momentum $p_t$. Its action in Hamiltonian
form is
\beq
S = \int ds \left( p_x \dot x + p_t \dot t - N H \right)
\label{2.1}
\eeq
where a dot denotes differentiation with respect to the
unphysical time parameter $s$. (Note that $t$ is physical time in this
section). $N$ is a Lagrange multiplier enforcing the
constraint
\beq
H = p_t + h = 0
\label{2.2}
\eeq
where $h$ is the usual Hamiltonian
\beq
h = \frac {p_x^2} { 2m} + V(x)
\eeq
Canonical quantization leads to the Schr\"odinger equation,
\beq
H |\psi \rangle  = \left( p_t + h \right) |\psi \rangle = 0
\label{2.3}
\eeq

We are ultimately interested in solutions to the constraint equation,
Eq.(\ref{2.3}), which are normalized in terms of an inner product
defined on spacetime (not just on space).
Following the general scheme for constructing the induced inner product,
we consider an enlarged Hilbert space ${\cal H}_x \otimes {\cal H}_t $,
where ${\cal H}_x$ is the usual Hilbert space of wave functions $\psi (x)$.
We may define states on this enlarged space of the form
\beq
| \Psi \rangle = \int dx dt \ | x \rangle \otimes | t \rangle \ \Psi (x,t)
\eeq
where $ | x \rangle $ and $ | t \rangle $ are eigenstates of
the operators $\hat x $ and $\hat t $ respectively.
We then consider eigenstates of
$H $,
\beq
H | \Psi_{\l} \rangle = \lambda | \Psi_{\l} \rangle
\label{2.4a}
\eeq
They are normalized using the auxiliary inner product
defined on ${\cal H}_x \otimes {\cal H}_t $,
\beq
\langle \Psi_{\l} | \Psi_{\l' }' \rangle_A = \int  dx dt \ \Psi_{\l }^* (x,t)
\Psi_{\l' }' (x,t)
\label{2.4}
\eeq
Since $H = p_t + h$,
the solutions to the eigenvalue equation may be written
\beq
\Psi_{\l } (x,t) = {  1 \over (2 \pi)^{\half} } \ e^{i \l t }
\psi (x,t)
\label{2.5}
\eeq
where $\psi (x,t)$ satisfies the Schr\"odinger equation, Eq.(\ref{2.3}).
It follows that
\beq
\langle \Psi_{\l} | \Psi_{\l' }' \rangle_A = { 1 \over 2 \pi}
\int dt \int dx \ e^{ - i \l t + i \l't }
\ \psi^* (x,t) \psi' (x,t)
\label{2.6}
\eeq
The integral contains within it the usual inner product
\beq
\langle \psi | \psi ' \rangle_S =
\int dx \ \psi^* (x,t) \psi' (x,t)
\label{2.7}
\eeq
This has the important property that it
is independent of time when the states obey
the Schr\"odinger equation, so the time integral may be done in
Eq.(\ref{2.6}), pulling down a delta function $\delta (\l - \l')$.
We thus obtain
\beq
\langle \Psi_{\l} | \Psi_{\l' }' \rangle_A
= \delta ( \l - \l') \langle \psi | \psi ' \rangle_S
\label{2.8}
\eeq
This means that the expected Schr\"odinger inner product
on surfaces of constant $t$ is fully compatible with the induced inner product
defined on the whole of spacetime.

We may now construct the decoherence functional for this system, for some interesting
physical questions. For a pure initial state, the decoherence functional is
\beq
D (\a, \a') = \langle \Psi_{\l'} | C_{\a'}^{\dag} C_{\a} | \Psi_{\l} \rangle
\label{2.9}
\eeq
It is constructed using the induced inner product and the class operators $C_{\a}$
must commute with the constraint, $H$.

\subsection{A Useful Mathematical Result}

The induced inner produce we are using here can be quite cumbersome and
it is useful to prove a simple mathematical result. The decoherence functional
is an expression of the general form
\beq
\langle \Psi_{\l'} | A | \Psi_{\l} \rangle
\label{2.10}
\eeq
where $A$ commutes with the constraint, and it is useful to show how this expression
reduces to a simpler expression on the original Hilbert space ${\cal H}_x$.
In all the expressions we are interested in, $A$ will commute with $\hat t$, so $A$
has the form
\beq
A = \int_{-\infty}^{\infty} dt \ B (t) \ \otimes \ | t \rangle \langle t |
\label{2.11}
\eeq
where $B(t)$ acts on ${\cal H}_x$ only. Using the fact that $[A,H] = 0$, it is
straightforward to deduce that
\beq
B(t) = e^{ - i h t} B (0) e^{ i h t}
\label{2.12}
\eeq
(where note that the signs in the exponents are not the ones associated
with Heisenberg picture evolution).
We now have
\bea
\langle \Psi_{\l'} | A | \Psi_{\l} \rangle
&=& \int dt \int dx' dx \ \Psi_{\l'}^* (x',t) \langle x' |
B(t) | x \rangle \ \Psi_{\l} (x,t)
\nonumber \\
&=& { 1 \over 2 \pi}
\int dt  \ e^{ - i \l t + i \l't } \int dx' dx \ \psi^* (x',t) \ \langle x' |
e^{ - i h t} B (0) e^{ i h t} | x \rangle \ \psi (x,t)
\label{2.13}
\eea
Now noting that
\beq
\int dx \ e^{ i h t} | x \rangle \ \psi (x,t) = \int dx  \ | x \rangle \ \psi (x,0) = | \psi \rangle
\label{2.14}
\eeq
we finally obtain
\beq
\langle \Psi_{\l'} | A | \Psi_{\l} \rangle = \delta (\l - \l') \ \langle \psi | B(0) | \psi \rangle
\label{2.15}
\eeq
The expression on the left is in terms of the auxiliary inner product on
${\cal H}_x \otimes {\cal H}_t $. The inner product
on the right is the usual one on ${\cal H}_x $. Hence expressions of the form
Eq.(\ref{2.10}) are readily evaluated once one has read off $B(t)$ in Eq.(\ref{2.12}).
(Note these expressions specifically refer to the parametrized non-relativistic
particle with Hamiltonian constraint Eq.(\ref{2.2}) -- they are not
valid for parametrized systems which are quadratic in all the momenta).

\subsection{Probabilities on Surfaces of Constant Time}

Now we consider the simple question, what is the probability of finding the
particle in a range $\Delta$ of the $x$-axis at time $t_0$? We of course expect
the standard answer
\beq
p_{\Delta} = \int_{\Delta} dx \ | \psi (x,t_0) |^2
\label{2.16}
\eeq
but it is important to see how this arises in a consistent quantization of the
parametrized particle.

We assert that the appropriate class operator corresponding to this question is
\beq
C_{\Delta,t_0} = \int_{-\infty}^{\infty} ds \ \delta ( \hat t (s) - t_0 )
\ f_{\Delta} ( \hat x (s) )
\label{2.17}
\eeq
where
\bea
\hat t(s) &=& e^{ i H s} \hat t e^{ - i H s} = t + s
\\
\hat x (s) &=& e^{ i H s } \hat x e^{ - i H s }
\label{2.18}
\eea
and $ f_{\Delta} (x)$ is a window function on the range $\Delta$. The class operator
clearly commutes with both $H$ and $\hat t$. Classically, for the free particle,
the class operator corresponds to the expression
\beq
C_{\Delta,t_0} = f_{\Delta} ( x - \frac {p} {m} ( t -t_0) )
\label{2.19}
\eeq
which clearly has the right properties: it is equal to $1$ for classical
trajectories which cross $t=t_0$ in $\Delta$ and zero otherwise. Returning to the
quantum case, one can see that
\beq
C_{\Delta,t_0} = \int_{-\infty}^{\infty} dt \ e^{ - i h (t-t_0)} f_{\Delta} (\hat x)
e^{ i h (t - t_0) } \ \otimes | t \rangle \langle t |
\label{2.20}
\eeq
This is of the form Eq.(\ref{2.11}) from which we read off
\beq
B(0) = e^{ i h t_0} f_{\Delta} (\hat x) e^{ - i h t_0 }
\label{2.21}
\eeq
Now a crucial simplification. Since $f_\Delta$ is a window function, $B(0)$ is in fact
a projection operator, and therefore the class operator $C_{\Delta, t_0}$ is also
a projection operator. This means that decoherence is automatic, between histories
characterized by $C_{\Delta,t_0}$ and $ 1 - C_{\Delta,t_0}$ and we may immediately
assign the probability
\beq
p_{\Delta} = \langle \Psi_{\l'} | C_{\Delta,t_0} | \Psi_{\l} \rangle
\label{2.22}
\eeq
Using Eq.({\ref{2.15}) and dropping the $\delta$-function, this becomes
\beq
p_{\Delta} = \langle \psi | e^{ i h t_0} f_{\Delta} (\hat x) e^{ - i h t_0 } | \psi \rangle
\label{2.23}
\eeq
which agrees exactly with the expected result Eq.(\ref{2.16}).
Note also that since the class operator is a projection operator in this case,
the decoherent histories analysis agrees exactly with the evolving constants
approach.

\subsection{Probabilities for Spacetime Regions}

We now consider a more challenging question which is to consider probabilities for regions
of spacetime. In particular, we pose the following question: given the initial wave function
$\psi (x)$ at $t=0$, what is the probability that the particle is found in the region
$x<0$ in the time interval $ [ 0 , \tau ]$? This has been analyzed previously in the decoherent
histories approach with the following results \cite{YaT}. (See also Refs.\cite{HaZa,Har4}).
The decoherence functional is
\beq
D(\a, \a') = \langle \psi | g_{\a'}^{\dag} g_{\a} | \psi \rangle
\label{2.24}
\eeq
in the usual inner product. (Here we use $g_{\a}$ to denote class operators to
avoid confusion with the class operators $C_{\a}$ defined on the enlarged space.)
There are two class operators. First, there is the class
operator for remaining in $ x> 0 $ during the time interval $[0,\tau]$, and this is
the restricted propagator $ g_r (\tau, 0 )$.
The other class operator may be written
\beq
g_c (\tau, 0 ) = e^{- i h \tau} - g_r (\tau, 0 )
\label{2.26}
\eeq
and corresponds to the situation in which the particle enters $ x<0$ at some time
during $[0,\tau]$. (It may also be expressed in terms of the path decomposition expansion
Eq.(\ref{pdx}), but we will not need this here. Note also that these class operators
reduce to the unitary operator $e^{- i h \tau}$ when the restrictions are removed).
The histories are generally not decoherent,
except for very special initial states, and the resultant probabilities are somewhat trivial \cite{YaT}.
However, our aim here is to show how the decoherence functional for this model is recovered
from a quantization of the parameterized particle as a constrained system on an enlarged
Hilbert space, in which its spacetime character is most transparent.

We take the decoherence functional Eq.(\ref{2.9}) and seek a class operator commuting with
$H = p_t + h $
corresponding to the statement that the particle never enters the region $x<0$
during the time interval $ [0,\tau]$. We denote this region $\Delta$ and we use
$\bar \Delta$ to denote the region outside $ \Delta$. In the geuninely spacetime point of
view used here, we may introduce projection operators onto the spacetime regions
$\Delta $ and $\bar \Delta$.
The projection onto $\Delta $ is
\beq
P = \theta ( \tau - \hat t ) \theta ( \hat t) \theta ( - \hat x)
\label{2.27}
\eeq
and the projection onto $\bar \Delta $ is conveniently written
\bea
\bar P &=& \theta ( - \hat t ) + \theta ( \tau - \hat t ) \theta ( \hat t) \theta ( \hat x)
+ \theta ( \hat t - \tau)
\\
&=& \int_{-\infty}^{\infty} dt\  \Upsilon (t, \hat x ) \otimes | t \rangle \langle t |
\label{2.28}
\eea
where $ \Upsilon (t, \hat x)$ is an operator on ${\cal H}_x$, equal to $\theta (\hat x)$
for $ 0 \le t \le \tau $ and equal to the identity otherwise.
The class operator for remaining in $\bar \Delta$ (that is, never entering the region $\Delta$)
is of the form
\beq
C_r = \prod_{s = \infty}^{\infty} \bar P (s)
\label{2.29}
\eeq
This example is sufficiently simple that we can work directly with the infinite
product (time-ordered) without encountering difficulties. We have, from Eq.(\ref{2.28}),
\bea
C_r &=& \int_{-\infty}^{\infty} dt\ \prod_{s = \infty}^{\infty}  \Upsilon (t+s, \hat x (s))
\otimes | t \rangle \langle t |
\nonumber \\
&=& \int_{-\infty}^{\infty} dt\ \prod_{s = -t}^{\tau - t} \ \theta ( \hat x (s) )
\otimes | t \rangle \langle t |
\label{2.30}
\eea
From the definition of the restricted propagator, Eq.(\ref{res}), we see that
\beq
C_r = \int_{-\infty}^{\infty} dt\ e^{ i h ( \tau - t )} g_r (\tau, 0) e^{ i h t}
\otimes | t \rangle \langle t |
\label{2.31}
\eeq
This commutes with the constraint $H$ and is of the desired from Eq.(\ref{2.11}) with
\beq
B(0) = e^{ i h \tau} g_r (\tau, 0)
\label{2.32}
\eeq
We are also interested in the quantity
\beq
C^{\dag}_r C_r = \int_{-\infty}^{\infty} dt \ e^{ - i h t} g_r (\tau, 0 )^{\dag}
g_r (\tau,0) e^{ i h t}
\otimes | t \rangle \langle t |
\label{2.33}
\eeq
which is also of the form Eq.(\ref{2.11}) with
\beq
B(0)  =  g_r (\tau, 0 )^{\dag} g_r (\tau,0)
\label{2.34}
\eeq
From these objects one can also construct the class operator for crossing,
\beq
C_c = 1 - C_r
\label{2.35}
\eeq
and related objects such as $C_c^{\dag} C_r $. It is now easy to see, using
Eq.(\ref{2.15}), that we readily obtain the known form of the decoherence
functional for this system. For example,
\beq
\langle \Psi_{\l'} | C^{\dag}_r C_r | \Psi_{\l} \rangle
= \delta ( \l' - \l) \langle \psi | g_r (\tau, 0 )^{\dag} g_r (\tau,0) | \psi \rangle
\label{2.36}
\eeq
which, via the induced inner product prescription,
agrees with the known result Eq.(\ref{2.24}).

These results show that our proposal for class operators passes the important test of the quantization of
the non-relativistic particle in parametrized form. Furthermore, there is the added feature
that it shows how spacetime questions in non-relativistic quantum mechanics may be
expressed in a genuinely spacetime form, since the decoherence functional and
probabilities (such as the left-hand side of Eq.(\ref{2.36})) may be expressed in terms
of an inner product and operators defined on spacetime.

\section{A Simple One-Dimensional Example}

The parametrized non-relativistic particle has the very special simplifying feature that
the Hamiltonian is linear in one of the momenta. This is not the case in general.
We therefore now consider some examples with a Hamiltonian quadratic in the momenta.
We first consider a simple one-dimensional example involving the
free particle. It is trivial in itself (except
to show that the class operators can be easily calculated), but
readily extends to higher dimension and
has important implications for the relativistic particle considered later.

\subsection{Energy Eigenstates}

We first consider normalization of the energy eigenstates. We have
\beq
H | \psi \rangle = E | \psi \rangle
\eeq
where $H = p^2 / 2m$. For each $E$ there are two solutions which
are conveniently written,
\beq
\psi^{\pm}_E (x) = \left( \frac{m } {2 E} \right)^{1/4}
\frac {e^{ \pm i | k | x }} { ( 2 \pi )^{1/2} }
\eeq
where $ | k | = \sqrt{2m E} $.
In the usual inner product,
\beq
\langle \psi_1 | \psi_2 \rangle = \int dx \ \psi_1^* (x) \psi_2 (x)
\eeq
we have
\beq
\langle \psi^{\pm}_E | \psi^{\pm}_{E'} \rangle = \delta (E - E')
\eeq
and
\beq
\langle \psi^{\pm}_E | \psi^{\mp}_{E'} \rangle = 0
\eeq
In the induced inner product prescription we therefore drop the $\delta$-function
and take the induced inner product between two eigenstates with the same $E$ to
be
\beq
\langle \psi^{\pm}_E | \psi^{\pm}_{E} \rangle_I = 1
\eeq
and
\beq
\langle \psi^{\pm}_E | \psi^{\mp}_{E} \rangle_I = 0
\eeq

\subsection{Class Operators for Crossing or Not Crossing the Origin}

Given these preliminaries,
we now consider the following simple question. Given that the system is in an
energy eigenstate, what is the probability that
the particle crosses or never crosses $x=0$, irrespective of time?
This is most easily handled by considering the class operator $C_r $ describing the situation
in which the particle is always in $x>0$ or $x<0$. The class operator $C_c$ for crossing $x=0$
is then given by
\beq
C_c = 1 - C_r
\label{4.1}
\eeq

Let $P$ be the projector onto the positive $x$-axis,
\beq
P = \int_0^\infty dx \ | x \rangle \langle x | = \theta (\hat x )
\label{4.2}
\eeq
(we use the hat symbol for operators only where clarity demands it).
Then
\beq
\bar P = 1 - P = \theta ( - \hat x)
\label{4.3}
\eeq
is the projector onto the negative $x$-axis. The class operator $C_r$ for
remaining in $x>0$ or $x<0$ is then given by, in a loose notation,
\beq
C_r = \prod_t P (t) + \prod_t \bar P (t) = C_r^+ + C_r^-
\label{4.4}
\eeq
where $+$ and $-$ denote the terms projecting onto the positive and negative $x$-axis
respectively.
This expression is defined more formally in terms of restricted propagators,
as in Eq.(\ref{3.18}).

This situation is sufficiently simple for the method of images to work. The restricted
propagator for the region $x> 0 $ is therefore
\beq
g_r^+ (x^{\pp},t | x', 0 ) = \theta (x^{\pp} ) \theta (x')
\left[ g(x^{\pp},t|x',0) -  g(x^{\pp},t|-x',0)\right]
\label{4.5}
\eeq
where
\beq
g(x^{\pp},t|x',0) = \left( \frac {m} {2 \pi i t} \right)^{1/2}
\exp \left(  \frac {i m} {2 t} (x^{\pp} - x')^2 \right)
\label{4.5b}
\eeq
is the free particle propagator.
The restricted propagator
may be usefully thought of as sum of two parts. The first term (in a semiclassical view)
corresponds to the direct path from $x'$ to $x^{\pp}$. The second term is usually
thought of as propagation from the image point $-x'$. However, it
may also be thought
of as corresponding to the path which again starts at $x'$ but is reflected off
$x=0$ before arriving at $x^{\pp}$. Differently put, the classical limit of a
system described by a restricted propagator is one in which the Hamiltonian includes
an infinite potential barrier at the boundary of the region in question, so its
classical trajectories include paths which reflect off the boundary.
These points will be important in the interpretation of the quantum results.

It is in fact quite useful to
rewrite this in an operator form using the projection operators
introduced above. We also introduce
the reflection operator
\beq
R = \int_{-\infty}^{\infty} dx \ | x \rangle \langle - x |
= \int_{-\infty}^{\infty} dp \ | p \rangle \langle - p |
\label{4.6}
\eeq
and note that $ [ H, R ] = 0 $ and $R^2 = 1$.
The restricted propagator in $x>0$ may then be written in operator form
as
\beq
g_r^+ (t,0) = P e^{- i H t } ( 1 - R) P
\label{4.7}
\eeq
From Eq.(\ref{3.18}), the desired class operator $C_r^+$ is
\beq
C_r^+ = \lim_{t^{\pp} \rightarrow \infty, t' \rightarrow -\infty} \ P(t^{\pp}) \ (1 - R) \ P (t')
\label{4.8}
\eeq
Now we need to take the infinite time limit in $P(t^{\pp})$ and $P(t')$.  We have
\beq
P (t) = \theta (\hat x_t ) = \theta ( \hat x + \frac{ \hat p t } {m} )
\label{4.9}
\eeq
Naively, for very large positive or negative $t$ the momentum term dominates, and we
get
\bea
\lim_{t^{\pp} \rightarrow \infty} P (t^{\pp} ) &=& \theta (\hat p)
\label{4.9b} \\
\lim_{t' \rightarrow - \infty} P( t') &=& \theta ( - \hat p )
\label{4.10}
\eea
However, it is important to verify this result more carefully, by sandwiching it
between an arbitrary pair of states. We have, using Eq.(\ref{4.5b}),
\bea
\langle \psi_1 | P(t) | \psi_2 \rangle
&=&  \frac {m} {2 \pi  t} \int_0^{\infty} dx \int_{-\infty}^{\infty} dy_1 \int_{-\infty}^{\infty} dy_2
\ \psi_1^* (y_1)  \ \psi_2 (y_2)
\nonumber \\& \times & \exp \left( - \frac {im} {2t} (y_1-x)^2 + \frac {im} {2t} ( x - y_2)^2
\right)
\eea
Defining $p = m x / t $, this becomes
\bea
\langle \psi_1 | P(t) | \psi_2 \rangle
&=&  \frac {1} {2 \pi  } \int_0^{\infty} dp \int_{-\infty}^{\infty} dy_1 \int_{-\infty}^{\infty} dy_2
\ \psi_1^* (y_1)  \ \psi_2 (y_2)
\nonumber \\& \times & \exp \left( i p (y_1- y_2) - \frac {im} {2t} ( y_1^2 - y_2^2)
\right)
\eea
Now, taking the limit $t \rightarrow \infty$, the last term in the exponent drops out,
and the integrals over $y_1$ and $y_2$ produce the Fourier transformed wave functions,
so we have
\bea
\lim_{t \rightarrow \infty}
\langle \psi_1 | P(t) | \psi_2 \rangle
&=& \int_0^{\infty} dp  \ \tilde \psi_1^* (p) \ \tilde \psi_2 (p)
\nonumber \\
&=& \langle \psi_1 | \theta ( \hat p ) | \psi_2 \rangle
\label{4.10b}
\eea
This confirms the naive result Eq.(\ref{4.9b}).

We now have
\bea
C_r^+ &=& \theta ( \hat p ) ( 1 - R ) \theta ( - \hat p )
\nonumber \\
&=& - \theta ( \hat p ) \ R \ \theta ( - \hat p )
\label{4.11}
\eea
which may also be written
\beq
C_r^+ = - \theta ( \hat p ) R = - R \theta ( - \hat p )
\label{4.11b}
\eeq
Note that the first part of the restricted propagator (involving
the direct path from initial to final point) drops out, leaving
only the part corresponding to the reflected paths.
In the momentum representation, this is
\beq
\langle k_2 | C_r^+ | k_1 \rangle
= - \theta (k_2 ) \delta (k_2 + k_1 )
\label{4.12}
\eeq

Similarly, the class operator $C_r^-$ is constructed using the restricted
propagator in $ x < 0$,
\beq
g_r^- (x^{\pp},t | x', 0 ) = \theta (-x^{\pp} ) \theta (-x')
\left[ g(x^{\pp},t|x',0) -  g(x^{\pp},t|-x',0)\right]
\label{4.13}
\eeq
and one readily finds that
\beq
C_r^- = - \theta ( -\hat p ) R = - R \theta (  \hat p )
\label{4.14}
\eeq
Combining the $+$ and $-$ parts we have
\beq
C_r =  - \theta ( \hat p ) R - \theta ( -\hat p ) R = - R
\label{4.15}
\eeq
a very simple expression. The class operator for crossing the surface is then
\beq
C_c = 1 - C_r = 1 + R
\label{4.16}
\eeq
As expected, all the above class operators, $C_c, C_r, C_r^+ $ and $C_r^-$ commute
with $H$.

\subsection{Decoherence Functional and Probabilities}

We may now compute the decoherence functional and the probabilities. The off-diagonal
part of the decoherence functional is
\beq
D (r,c) = {\rm Tr} \left( C_r \rho C_c^{\dag} \right)
\label{4.17}
\eeq
computed in the induced inner product.
We take a pure initial state which is an eigenstate of $H$.
Noting that
\beq
C_c^{\dag} C_r = - (1+R) R = - (R+1)
\label{4.18}
\eeq
we have
\beq
D(r,c) = -\langle \Psi_E | (1 + R ) | \Psi_E \rangle_I
\label{4.19}
\eeq
where $ | \Psi_E \rangle $ is an energy eigenstate and the
induced inner product is used. Using the formula Eq.(\ref{1.12c})
for the induced inner product, this means that
there is decoherence only for states satisfying
\beq
\langle \phi | \delta (H - E) (1 + R ) | \phi \rangle = 0
\label{4.20}
\eeq
where $ | \Psi_E \rangle = \delta (H - E) | \phi \rangle $.
(This expression is in fact real, so there is no difference between consistency
and decoherence). The operator $(1+R)$ produces the symmetric part of
$ | \phi \rangle$. The decoherence condition therefore implies that the
fiducial wave function must be antisymmetric,
\beq
\phi (-x) = - \phi ( x)
\label{4.21}
\eeq
This means that $ \Psi_E (x)$ must also be antisymmetric (using the fact
that $R$ commutes with $H$).
For such wave functions, probabilities are defined and we get
\beq
p_r = \langle \Psi_E | C_r^{\dag} C_r | \Psi_E \rangle_I = 1
\label{4.22}
\eeq
for the probability for not crossing
since $ C_r^{\dag} C_r  = R^2 = 1 $. Similarly, the probability
for crossing is zero.

It is useful to try and understand this result in terms of classical paths.
Recall that firstly, only entire infinite classical paths are reparmetrization
invariant and secondly, that the restricted propagator corresponds to a classical
situation in which there is an infinite barrier at $x=0$.
Classically, therefore, this result in some sense corresponds to classical trajectories
which remain in $x>0$ or $x<0$ by bouncing off $x=0$.

This result is not very physically enlightening but it is very similar
to the result obtained in the decoherent histories analysis of the
crossing time problem in non-relativistic quantum mechanics. There, one looks
for probabilities that, given an initial state, the particle will cross $x=0$
during the time interval $[0, \tau]$. One finds that only antisymmetric wave
functions give consistency and the crossing probability is zero \cite{YaT}.
More physically intuitive results in the crossing time problem are obtained|
when there is a decoherence mechanism in place \cite{HaZa}. We expect
that to be the case here, too, but this will explored in another paper \cite{HaWa}.

\subsection{A More Detailed Look at the Crossing Class Operator}

It is useful to give an alternative derivation of the crossing class operator
Eq.(\ref{4.16}) using the path decomposition expansion, Eq.(\ref{pdx}).
This is partly a consistency check but it
will also give some insight into the form of the result.

We first consider the fixed time crossing propagator, taking into account
the possibility of crossing the origin in either direction. Applying
Eq.(\ref{pdx}) together with the restricted propagators Eqs.(\ref{4.5}), (\ref{4.13}),
this is
\bea
g_c (x^{\pp}, t^{\pp} | x', t') &=& \lim_{x \rightarrow 0-}
\int_{t'}^{t^{\pp}} dt \ g (x^{\pp},t^{\pp}|x,t)
\ \frac {i} {2m} \frac {\partial} {\partial x} g_r^+ (x,t | x', t')
\\ \nonumber
&-& \lim_{x \rightarrow 0+}
\int_{t'}^{t^{\pp}} dt \ g (x^{\pp},t^{\pp}|x,t)
\ \frac {i} {2m} \frac {\partial} {\partial x} g_r^- (x,t | x', t')
\eea
where the relative minus sign is because of the definition of the normal
in Eq.(\ref{pdx}). Now, using Eqs.(\ref{4.5}), (\ref{4.13}), we have
\bea
\lim_{x \rightarrow 0-} \ \frac {\partial} {\partial x} g_r^- (x,t | x', t')
&=& 2 \theta (-x') \ \frac {\partial} {\partial x} g (x,t | x', t') \big|_{x=0}
\\
\lim_{x \rightarrow 0+} \ \frac {\partial} {\partial x} g_r^+ (x,t | x', t')
&=& 2 \theta (x')\  \frac {\partial} {\partial x} g (x,t | x', t') \big|_{x=0}
\eea
We therefore have
\beq
g_c (x^{\pp}, t^{\pp} | x', t') =  \lim_{x \rightarrow 0}
\int_{t'}^{t^{\pp}} dt \ g (x^{\pp},t^{\pp}|x,t)
\ \frac {i} {m} \frac {\partial} {\partial x} g (x,t | x', t') \ \epsilon ( x')
\eeq
where $\epsilon (x) $ is the signum function. This is conveniently written
in operator form as
\beq
g_c (t^{\pp},t') =  \frac {1} {m} \int_{t'}^{t^{\pp}} dt \ e^{ - i H (t^{\pp} - t)}
\ \delta (\hat x ) \ \hat p \ e^{ - i H (t-t')} \ \epsilon (  - \hat x)
\eeq
The desired crossing class operator is now given by
\bea
C_c &=& \lim_{t^{\pp} \rightarrow \infty, t' \rightarrow - \infty}
e^{ + i H t^{\pp} } g_c (t^{\pp},t')\  e^{- i H t'}
\nonumber \\
&=& \frac {1} {m} \int_{-\infty}^{\infty} dt  \ \delta ( \hat x_t) \ \ | \hat p |
\label{4.25}
\eea
where we have used the fact that $ \epsilon (- \hat x_t) \rightarrow \epsilon ( \hat p)$
as $ t \rightarrow - \infty$ and $ | \hat p | = \hat p \ \epsilon (\hat p)$.

Eq.(\ref{4.25}) is the desired expression and shows very clearly that the class operator
for crossing $x=0$ involves some kind of flux at $x=0$. Classically, it is easy
to see that this expression is equal to $1$ for all classical paths, except those
for which $p=0$, in which case it is zero. Hence it clearly encapsulates the classical notion
of surface crossing. One might argue that classical states with $p=0$ are a set of
measure zero so may be safely neglected. However, the $p=0$ states seem to be crucial
to understand the quantum case.

As an operator expression, Eq.(\ref{4.25}) may be evaluated by sandwiching it between two momentum
states:
\bea
\langle p^{\pp} | C_c | p' \rangle
&=& \frac {1} {m} \int_{-\infty}^{\infty} dt \ \exp \left(  \frac {it} {2m} ( {p^{\pp}}^2 -{p'}^2)
\right)
\ \langle p^{\pp} | \delta ( \hat x ) | p' \rangle \ | p ' |
\nonumber \\
&=& 2 | p' | \ \delta (  {p^{\pp}}^2 -{p'}^2)
\nonumber \\
&=& \delta ( p^{\pp} - p') + \delta ( p^{\pp} + p')
\eea
It follows that
\beq
C_c = 1 + R
\label{4.26}
\eeq
the expected result.

How are we to understand this result? Classically, we noted that all trajectories
with $p \ne 0 $ cross the origin. The result Eq.(\ref{4.26}) ought therefore to be the
quantum implementation of this idea. The key thing is that the operator
$1+R$ is zero when acting on states which are antisymmetric in $x$ about the origin. Such
states are also antisymmetric in $p$ so clearly vanish at $p=0$. Hence the crossing
class operator annihilates a class of states with $p=0$ and in this sense implements
the classical notion of crossing.

Of course, there are many inequivalent ways to turn classical expressions
into quantum operators, and one could imagine that a quantization procedure may
exist which consistently drops the $p=0$ states before quantization.
This would avoid the difficulties of interpretation
with reflected paths encountered earlier. However, this does not appear to be the
case in the present quantization method.

We also remark that, as noted earlier, the evolving constants method also encounters
difficulties with $p=0$ states because of the $1/p$ factors arising in the evolving
constants operators.

\section{The Relativistic Particle}

The result of the previous section is readily extended to the relativistic
particle in $3+1$ dimensions described by the constraint equation
\beq
H | \psi \rangle = \left( p_0^2 - {\bf p}^2 \right) | \psi \rangle = 0
\label{5.1}
\eeq
We consider this in order to compare with
a previous attempt to define class operators compatible with the constraint \cite{HaTh1,HaMa}.

Consider the following question. What is the probability
that a free relativistic particle never crosses the spacelike surface $x^0 = 0$?
Classically, this probability must be zero, because every (timelike) classical trajectory
crosses any spacelike surface somewhere, since the classical trajectories
are just straight lines.
In Ref.\cite{HaTh1}, a (somewhat ad hoc)
proposal was made to define class operators compatible with the constraint which
coincide with this classical intuition. A quantum-mechanical probability of
zero was thus obtained for not crossing the surface.

The question is readily addressed in the present approach using
an elementary extension of the results of the
one-dimensional model in the previous section. The class operator
for not crossing $x^0 = 0$ is of the form Eq.(\ref{4.4}) where the Hamiltonian
is as in Eq.(\ref{5.1})
and the projectors $P$ and $\bar P$ project onto the regions $x^0 >  0 $
and $x^0 < 0 $ respectively. One easily finds that decoherence
is only possible for states antisymmetric about $x^0 = 0$ and the probability
for not crossing is equal to $1$. It is therefore the exactly opposite result
to that obtained in Ref.\cite{HaTh1}

What is the origin of the difference in results and which is the ``correct'' one?
The key point is that in the present approach, the restricted propagators
involved in the construction of the class operators,
such as Eq.(\ref{4.5}) involve two types of paths in a semiclassical picture,
the direct paths and the reflected paths. As noted above (after Eq.(\ref{4.11b})),
the part of the propagators corresponding to the direct paths drops out, so the present
approach consists entirely of the contribution from the reflected paths. The
classical intuition of Ref.\cite{HaTh1} was tacitly based on the direct paths, so the
result is completely different. Since the reflected paths capture the important
notion of the reflection of wave packets from a barrier,
it is appropriate to take the present approach as the definitive one
if we are to stay true to quantum-mechanical ideas.

In the closely related context of the arrival time problem, it has however been
shown that, in the presence of a decohering environment, the effect of the reflected
paths becomes less significant, and this is how classical intuition may become restored \cite{HaZa}.
See also Ref.\cite{Whe} for further relevant considerations of the relativistic particle.

This brief discussion of the relativistic particle indicates that the present proposal
for class operators is not in fact a more developed statement of the ad hoc approach of
Refs.\cite{HaMa,HaTh1,HaTh2}, but a different proposal altogether.


\section{The Free Particle in Two Dimensions}

In two dimensions more interesting questions are possible. However,
the general difficulty one expects to encounter for most questions
is decoherence. In the absence of an environment, most situations
will not have decoherent sets of histories. In the
one-dimensional example, the lack of decoherence for general states
is largely due to the feature of reflection at the boundaries of
the regions of interest. It is therefore of interest to consider
situations where this reflection will not happen.

In two (and more) dimensions, we have the possibility of eigenstates
of $H$ which are rotationally symmetric about the origin and may be
thought of as superpositions of wave packets moving radially.
It therefore seems plausible that if we consider questions concerning
regions whose boundaries lie along radial lines, there will be little
or no possibility of crossing or reflection (since the wave function has no
flux across the boundary). For example, given a rotationally symmetric state,
we could
ask for the probability that the particle is found in a wedge-shaped
region emanating from the origin. This question bears some resemblance
to Mott's calculation of alpha-ray tracks \cite{Mott}. He showed,
using a series of model detectors, why an outgoing spherical wave
produces a straight line track in a detector.

\subsection{Particle in The First Quadrant}

We first consider the following question. Given that the system
is in an energy eigenstate, what is the probability that the particle will
always be in the region, $x>0, y>0$?
We denote this region $\Delta$ and
use the method of images to construct the restricted propagator
for propagation in $\Delta $. It is
\bea
g_{\Delta} (x^{\pp}, y^{\pp}, t | x', y', 0 ) & = &
 \theta ( x^{\pp}) \theta (  y^{\pp} )
\ \theta ( x' ) \theta ( y' )
\nonumber \\ & \times &
[ g (x^{\pp}, y^{\pp}, t | x', y', 0 )
- g (x^{\pp}, y^{\pp}, t | -x', y', 0 )
\nonumber \\
 &-& g (x^{\pp}, y^{\pp}, t | x', -y', 0 )
+ g (x^{\pp}, y^{\pp}, t | -x', -y', 0 ) ]
\eea
where $  g (x^{\pp}, y^{\pp}, t | x', y', 0 )$ is the free particle propagator
in two dimensions.
As in the previous example, an operator form is useful. We introduce two reflection
operators
\bea
R_x &=& \int dx dy \ | x,y \rangle \langle -x,y |
\\
R_y &=& \int dx dy \ | x,y \rangle \langle x,-y |
\eea
The class operator for histories which are always in this region is therefore
\beq
C_{\Delta} = \lim_{t^{\pp} \rightarrow \infty, t' \rightarrow -\infty} \ P(t^{\pp}) \
\left( 1 - R_x - R_y + R_x R_y \right) \ P (t')
\label{FP3}
\eeq
where
\beq
P(t) = \theta ( \hat x_t ) \theta (  \hat y_t )
\eeq
Following the same method as in the previous example, we therefore get
\beq
C_{\Delta} =  \theta ( \hat p_x ) \theta (  \hat p_y )
\ \left( 1 - R_x \right) \left( 1- R_y \right)
\   \theta ( -\hat p_x ) \theta ( - \hat p_y )
\eeq
Using the properties of the reflection operators this is easily seen to be,
\beq
C_{\Delta} = \theta ( \hat p_x ) \theta (  \hat p_y ) R_x R_y
\label{FP4}
\eeq
For a pure initial state, the off-diagonal part of the decoherence functional is
\bea
D(\bar \Delta , \Delta ) &=& \langle \psi | ( 1 - C_{\Delta})^{\dag} C_{\Delta} | \psi \rangle_I
\nonumber \\
&=& \langle \psi | (R_x R_y - 1)  \theta ( -\hat p_x ) \theta ( - \hat p_y ) | \psi \rangle_I
\eea
where $ | \psi \rangle $ are energy eigenstates and the induced inner product is used.
(In what follows we will not spell out the induced inner product calculation
since it is very similar to that in Section 6).
From this it is easy to see that we get decoherence if the state satisfies
\beq
R_x R_y | \psi \rangle = | \psi \rangle
\label{FP5}
\eeq
This condition is easily satisfied, as expected, by a rotationally symmetric state
and the probability for remaining in the region $\Delta$ is
\beq
p_{\Delta} = \langle \psi |\theta ( \hat p_x ) \theta (  \hat p_y ) | \psi \rangle_I
\label{FP6}
\eeq
For rotationally symmetric states this will be equal to $1/4$, the expected result.

Classically, this result corresponds to straight line trajectories radiating
from the origin in the positive quadrant. One can think of them as coming in from
infinity, bouncing at the origin and returning to infinity.

\subsection{Particle in the First and Third Quadrant}

Since the classical trajectories are straight lines, as a modification of the above
calculation, it seems reasonable to also consider the situation in which the region
$\Delta$ is the first and third quadrant. We can then ask whether it is possible to have a
situation which corresponds classically to a family of infinite straight lines passing
through the origin.

There is a subtlety, however, in that there are two similar but different class operators.
First, one can ask for the probability that the particle remains always in the
first quadrant $x>0, y>0$ or always in the third quadrant, $x<0, y<0$. It is easy
to see that the appropriate class operator has the loose form
\beq
C_{\Delta} = \prod_t P_1 (t) + \prod_t P_3 (t)
\label{FP7a}
\eeq
where $P_1 $ and $P_3$ are projectors onto the first and third quadrants,
respectively. Or more formally, this is
\beq
C_{\Delta} = \left[ \theta ( \hat p_x ) \theta (  \hat p_y )
+ \theta ( - \hat p_x ) \theta (  -\hat p_y ) \right]
R_x R_y
\label{FP7}
\eeq
We again get decoherence for states satisfying Eq.(\ref{FP5}) and the probability
associated with $\Delta $ is $1/2$.

Second, one can ask for the probability that the particle is in the first or third quadrant,
but in addition, has the possibility of following trajectories which lie in {\it both}
quadrants. (This possibility is not present in the class operator Eq.(\ref{FP7})).
The class operator for this second case has the loose form,
\beq
C_{\Delta} = \prod_t ( P_1(t) + P_3 (t) )
\eeq
which is clearly different from Eq.(\ref{FP7a}). By comparing with Eq.(\ref{FP3}),
it is easily see that
this class operator is
\bea
C_{\Delta} &=& \left[ \theta ( \hat p_x ) \theta (  \hat p_y )
+ \theta ( - \hat p_x ) \theta (  -\hat p_y ) \right]
( 1 - R_x - R_y + R_x R_y) \left[ \theta ( \hat p_x ) \theta (  \hat p_y )
+ \theta ( - \hat p_x ) \theta (  -\hat p_y ) \right]
\nonumber \\
&=&
\left[\theta ( \hat p_x ) \theta (  \hat p_y ) + \theta ( - \hat p_x ) \theta (  -\hat p_y )\right]
( 1 + R_x R_y )
\eea
It is easy to show that
\beq
C^{\dag}_{\Delta} C_{\Delta} = 2 C_{\Delta}
\eeq
from which it follows that the off-diagonal term in the decoherence functional is
\bea
D(\bar \Delta , \Delta ) &=& \langle \psi | ( 1 - C_{\Delta})^{\dag} C_{\Delta} | \psi \rangle_I
\nonumber \\
&=& -  \langle \psi | C_{\Delta} | \psi \rangle_I
\eea
The factor of $ (1 + R_x R_y ) $ in $C_\Delta$ means that there is only decoherence
for states satisfying
\beq
R_x R_y | \psi \rangle = - | \psi \rangle
\eeq
This is the opposite condition to the previous case and is not satisfied by the
rotationally symmetric wave function. Moreover, the off-diagonal part of the
decoherence functional is in fact proportional to the probability, so for
states satisfying the decoherence condition, the probability associated with
$\Delta$ is zero. There does not appear to be an obvious interpretation of this
result in terms of classical trajectories.

\subsection{Particle in a Wedge-Shaped Region}

We now consider the case of a more general region $\Delta$ consisting of
the wedge lying in the region $ 0 \le \phi \le \beta$ (in polar coordinates
$r,\phi$). We again ask for the probability that the particle is always
in the region $\Delta$.

Following the general scheme, we first require the time-dependent propagator
for the wedge region. For simplicity we restrict to the case where the angle
$\beta $ is $\beta =  \pi / b $, where $b$ is an integer. We also take $b$
to be even (which turns out to be simplest to deal with).
Then the restricted propagator
for the interior of the region is
\bea
g_{\beta} (x,y, t | x_0, y_0, 0 ) &=& f_{\beta} (x,y) \  f_{\beta} (x_0, y_0)
\nonumber \\
&=&
\sum_{n=0}^{b-1} \left[
g (r, 2 n \beta  + \phi, t | r_0, \phi_0, 0) -
g (r, 2 n \beta  - \phi, t | r_0, \phi_0, 0) \right]
\eea
where $g(r,\phi,t | r_0, \phi_0, 0 ) $ is the free particle propagator
in two dimensions in polar coordinates, and $f_{\beta} (x,y)$ is a characteristic
function equal to $1$ inside the wedge region and zero outside \cite{wedges}.
The desired class
operator is now
\beq
C_{\beta} = \lim_{t^{\pp} \rightarrow \infty, t' \rightarrow - \infty}
P(t^{\pp}) \ \sum_{n=0}^{b-1} \left[ R_n - K_n \right]
\ P(t')
\eeq
where the projector $P$ is $ f_{\beta} ( \hat x, \hat y) $ and we have introduced
the rotation operators
\bea
R_n &=& \int r dr d \phi \  | r, 2 n \beta + \phi \rangle \langle r, \phi |
\\
K_n &=& \int r dr d \phi \  | r, 2 n \beta -\phi \rangle \langle r, \phi |
\eea
Using the same method as in Eq.(\ref{4.10b}),
it may be shown that
\beq
\lim_{t \rightarrow \infty} P(t) = f_{\beta} (\hat p_x, \hat p_y)
\eeq
so we obtain
\beq
C_{\beta} =   f_{\beta} (\hat p_x, \hat p_y) \ \sum_{n=0}^{b-1} \left[ R_n - K_n \right] f_{\beta}
(- \hat p_x, -\hat p_y)
\eeq

Since $R_n$ rotates by an angle $2 n \beta$, we have
\beq
f_{\beta} (\hat p_x, \hat p_y) R_n f_{\beta} (-\hat p_x, -\hat p_y)
= f_{\beta} (\hat p_x, \hat p_y) f_{\beta} (- \hat p_x^n, -\hat p_y^n)
R_n^{-1}
\label{FP11}
\eeq
where $p_x^n$, $p_y^n$ denote the momenta rotated through angle
$2 n \beta$. Clearly, Eq.(\ref{FP11}) is zero unless $n = b/2$
(recall that $b$ is an even integer). Similarly, it is readily shown that
\beq
f_{\beta} (\hat p_x, \hat p_y) K_n f_{\beta} (-\hat p_x, -\hat p_y)
= 0
\eeq
We now have
\beq
C_{\beta} = f_{\beta} (\hat p_x, \hat p_y) R_{b/2}
\eeq
This agrees with Eq.(\ref{FP4}) when $ \beta =  \pi/ 2 $ as expected.

The off-diagonal term in the decoherence functional is
\bea
D( \bar \beta, \beta) &=&  \langle \psi | (1 - C_{\beta} )^{\dag} C_{\beta} | \psi \rangle_I
\nonumber \\
&=& \langle \psi | (R_{b/2} - 1) f ( -\hat p_x, -\hat p_y) | \psi \rangle_I
\eea
in the induced inner product, where $ | \psi \rangle $ are energy eigenstates.
This means that there is decoherence for states satisfying
\beq
R_{b/2} | \psi \rangle = | \psi \rangle
\eeq
This will indeed be satisfied for the rotationally symmetric state and the probability
then is
\bea
p_{\beta} &=&  \langle \psi | C_{\beta} | \psi \rangle_I
\nonumber \\
&=& \langle \psi | f_{\beta} (\hat p_x, \hat p_y) | \psi \rangle_I
\eea
By symmetry, we clearly have
\beq
p_{\beta} = \frac {\beta} {2 \pi}
\eeq
as expected.

To summarize this section, we have obtained some reasonable physical results
for rotationally symmetric wave functions and wedge-shaped regions, although
some simple extensions of these ideas run into some interpretational difficulties.
These examples uphold some of our intuitions, and it is gratifying that it is not
always necessary to employ an environment to get decoherence.

\section{Systems of Harmonic Oscillators}

The formalism so far concerned unbound systems, in which the (unphysical) time-parameter $t$
runs from $-\infty$ to $+ \infty$. It is however very different (and simpler)
for systems of harmonic oscillators, which are periodic in time.
In this section we consider the case of a $d$-dimensional simple harmonic
oscillator with Hamiltonian
\beq
H = \half \left( \p^2 + \x^2 \right)
\label{7.1}
\eeq
(See Ref.\cite{Rov1} for the evolving constants analysis of this system).
Much of the formalism will, however, be applicable to other systems periodic
in time. Systems described by the Hamiltonian Eq.(\ref{7.1}) will have period
$ 2 \pi$ so in the quantum theory we have
\beq
e^{i H (t + 2 \pi)} =  e^{ i H t}
\label{7.1a}
\eeq
An important class of observables for this system are of the form
\beq
A = \int_0^{2 \pi} dt \ B(t)
\label{7.1b}
\eeq
It is easy to show that $A$ commutes with $H$ using the fact that
the periodicity implies the property
\beq
\int_0^{2 \pi} dt \ B(t) = \int_{\tau}^{ 2 \pi + \tau} dt \ B(t)
\label{7.1c}
\eeq
for any $\tau$.
Note that
because the spectrum of $H$ is discrete, it is not necessary to use the induced inner product.

\subsection{Class Operators}

For the systems described by Hamiltonian Eq.(\ref{7.1}),
the natural modification of Eq.(\ref{3.13}),
the class operator for not entering the region $\Delta$, is
\beq
C_{\bar \Delta} = \prod_{t=0}^{2 \pi} \bar P (t)
\label{7.2}
\eeq
where $\bar P$ is the projector onto the region outside $\Delta$.
However, it is not hard to see that this does not in fact commute with $H$ (unless
the $P(t)$ all commute at different times). This is an operator ordering issue and is easily remedied
by defining the class operator to be instead
\beq
C_{\bar \Delta} =
\frac {1} {2 \pi} \int_0^{2 \pi} ds \prod_{t=s}^{s+ 2 \pi} \bar P (t)
\label{7.3}
\eeq
This is essentially a sum over all cyclic permutations of the operators
$\bar P(t)$ at different times, and now commutes with the Hamiltonian.

Following steps similar to those used in previous Sections, it is easily shown that
\beq
C_{\bar \Delta} = \frac {1} {2 \pi}
\int_0^{2 \pi} ds \ e^{ i H ( s + 2 \pi)} \ g_r( s+ 2 \pi, s)
\ e^{ - i H s}
\label{7.4}
\eeq
where $g_r$ is the restricted propagator. (Note that the restricted propagator
will not in general be periodic in time.)
Using Eq.(\ref{7.1a}) together with the fact
that $g_r (t,t')$ depends on time only through $(t-t')$, we have
\beq
C_{\bar \Delta} = \frac {1} {2 \pi}
\int_0^{2 \pi} ds \ e^{ i H  s } \ g_r( 2 \pi, 0)
\ e^{ - i H s}
\label{7.4a}
\eeq
This is of the form Eq.(\ref{7.1b}) so commutes with $H$, as expected.

The class operator $C_{\Delta} = 1 - C_{\bar \Delta} $ may be written,
\beq
C_{\Delta} = \frac {1} {2 \pi}
\int_0^{2 \pi} ds \ e^{ i H  s } \ g_c( 2 \pi, 0)
\ e^{ - i H s}
\label{7.5}
\eeq
where $g_c (2 \pi, 0 )$ is the crossing propagator, given by a sum over paths which enter
$\Delta$ at some time during the interval $[0, 2 \pi]$.

\subsection{Decoherence Functional and Probabilities}

We may now look at the decoherence functional and the probabilities. We choose a pure
initial state $ | \psi \rangle $ and, in keeping with the general approach, this state
is taken to be an eigenstate of the Hamiltonian
\beq
H | \psi \rangle = E | \psi \rangle
\label{7.13}
\eeq
The off-diagonal
term of the decoherence functional is
\beq
D (\Delta, \bar \Delta ) = \langle \psi | C_{\Delta}^{\dag} C_{\bar \Delta} | \psi \rangle
\label{7.14}
\eeq
Inserting the explicit
expressions, Eqs.(\ref{7.4a}), (\ref{7.5}), we have
\beq
D (\Delta, \bar \Delta )
= \int_0^{2 \pi} \frac{ ds_1 } {2 \pi}
\  \int_0^{2 \pi} \frac{ ds_2 } {2 \pi}
\ \langle \psi | e^{ i H s_1} g_c^{\dag} (2 \pi, 0 ) e^{  - i H (s_1 - s_2) } g_r (2 \pi, 0) e^{ - i H s_2}
| \psi \rangle
\label{7.15}
\eeq
Using Eq.(\ref{7.13}) this becomes
\beq
D (\Delta, \bar \Delta )
= \langle \psi | g_c^{\dag} (2 \pi, 0 ) \ P_E \  g_r (2 \pi, 0) | \psi \rangle
\label{7.16}
\eeq
where we have introduced the object
\beq
P_E =  \int_0^{2 \pi} \frac{ ds } {2 \pi} \ e^{ - i (H - E) s }
\label{7.17}
\eeq
Because the spectrum of $H$ is discrete, this is a projection operator, so satisfies
$P_E^2 = P_E$.

The decoherence functional will not be diagonal in general, although one simple
case in which it will is when the wave function is an eigenstate of $g_c (2 \pi, 0)$
or $g_r (2 \pi, 0) $.
We will exhibit such states below.

When the decoherence condition is satisfied, the probabilities associated with $\Delta$
and $\bar \Delta $ are easily shown to be
\bea
p_{\Delta} &=&  \langle \psi | g_c (2 \pi, 0) | \psi \rangle
\nonumber \\
p_{\bar \Delta} &=&  \langle \psi | g_r (2 \pi, 0)  | \psi \rangle
\label{7.18}
\eea

\subsection{Some Special States Exhibiting Approximate Decoherence}

We now introduce some states which exhibit approximate decoherence
and quasiclassical behaviour for this model. Consider first
the standard coherent states of the harmonic oscillator,
$ | \p, \x \ra $. They are preserved in form under
unitary evolution,
\beq
e^{ - i H t } | \p, \x \ra = e^{ - i t / 2}| \p_t, \x_t \ra
\label{7.21}
\eeq
where $ \p_t, \x_t $ are the classical solutions matching
$\p, \x $ at $t=0$, hence they are strongly peaked about
the classical path. There is a set of states which are
natural analogues of these states for the timeless models
considered here. They were referred to in Ref.\cite{HalY} as
``timeless coherent states'' and are defined by
\bea
| \phi_{\p \x} \rangle &=&  P_E | \p, \x \rangle
\nonumber \\
&=&  \int_0^{2 \pi} { dt \over 2 \pi } \ e^{ - i (H - E ) t }
\ | \p , \x \rangle
\nonumber \\
&=&   \int_0^{2 \pi} { dt \over 2 \pi } \ e^{i (E  - \frac{1}{2}) t }\ | \p_t, \x_t \rangle
\label{7.22}
\eea
They are clearly eigenstates of $H$ with eigenvalue $E$ and
are concentrated around the entire classical path with initial data
$ \p, \x $. They are not normalized to $1$ exactly, but if the initial
data satisfies $ E = \frac {1} {2} ( \p^2 + \x^2 ) $ then the coherent
states $ | \p, \x \rangle $ are approximate eigenstates of
$P_E$ and the timeless coherent states are then approximately normalized
to $1$. Further properties of these states are described in Ref.\cite{HalY}.

Now consider a timeless coherent state whose trajectory $\p_t, \x_t$ lies
entirely within the region $\bar \Delta$. This region could, for example,
be a large rectangular region in configuration space.
Or it could be a tube
following the classical trajectory but broadened out beyond the scale
of quantum fluctuations. In both these cases, if $\bar P$ is the projector onto the region
$\bar \Delta$, then we clearly have
\beq
\bar P | \phi_{\p \x} \rangle \ \approx \ | \phi_{\p \x } \rangle
\label{7.23}
\eeq
and
\beq
P | \phi_{\p \x} \rangle \ \approx \ 0
\eeq
We assert that with this choice of $\bar \Delta $,
the state $ | \phi_{\p \x} \rangle $ will give approximate decoherence.
There are two ways to see this.

First, from the (informal) expression Eq.(\ref{7.3}), the result Eq.(\ref{7.23})
together with the fact that the state is an eigenstate of $H$ imply that it is also
an approximate eigenstate of $\bar P(t)$, so will be an approximate eigenstates
of the class operator. This means there is approximate decoherence.

Second, and perhaps a little more rigorously, we use the expression
for the decoherence functional Eq.(\ref{7.16}). The important thing is to consider
the action of the restricted propagator $g_r (2 \pi, 0 )$ on the state
$ | \phi_{\p \x} \rangle $, which, via Eq.(\ref{7.22}), boils down to
its action on the coherent state $ | \p, \x \rangle $.
Restricted propagators are very difficult to calculate for arbitrary regions,
but their path integral form gives an intuitive picture
of their properties. It is
\beq
g_r ( \x^{\pp}, 2 \pi | \x', 0 )
= \int_{\bar \Delta} {\cal D} \x \exp \left( i \int_0^{2 \pi} dt
\left[ \half \dot \x^2 - \half \x^2 \right] \right)
\eeq
This is a sum over paths $\x (t)$ which remain always in the region $\bar \Delta$
and satisfying the end-point conditions $ \x (0) = \x'$, $\x (2 \pi) = \x^{\pp}$.
Suppose that the trajectory $ \p_t, \x_t $ of the coherent state $ | \p, \x \rangle$
remains entirely within the region $\bar \Delta $ (and does not approach the boundary).
Then, when this initial state is attached to the restricted propagator, 
the path integral will be dominated by the classical path with initial data $ \p, \x$.
The path integral will therefore be approximately the same as the unrestricted
path integral, which means that
\beq
g_r ( 2 \pi, 0 ) | \p, \x \rangle \approx | \p, \x \rangle
\eeq
It follows that there will be approximate decoherence and the probability for finding
the particle in the region $\bar \Delta$ is approximately $1$.

So for these specially chosen regions that entirely contain the trajectory of the
timeless coherent state we get approximate decoherence and the expected probabilities.
Note that these heuristic arguments only work for periodic systems in which certain states remain
coherent. For the systems considered
in earlier sections involving an infinite range of time, the spreading of wave packets
would render such heuristic arguments invalid.

For most other choices of $\Delta$, however, there is no decoherence and probabilities
cannot be assigned without a decoherence mechanism. This will be pursued elsewhere
\cite{HaWa}.


\section{Summary and Conclusions}

We have discussed the issues involved in defining class operators for the decoherent
histories analysis of reparametrization invariant systems and made a
specific proposal for such operators.
The class operators defined are based on certain reasonable classical expressions
and reduce to projection operators when everything commutes.
They commute with the Hamiltonian so fully respect reparametrization invariance.
They do not, however, exhibit the localization properties of their
classical counterparts. This is because there is an incompatibility between localization
and the constraint and in our definition we have made the choice that the constraint should
take precedence.

We compared with the evolving constants approach and noted that the difference between
that approach and the present one concerned the different ways in which equivalent
classical expressions are turned into quantum operators.

We showed that our class operators gave the correct and expected results when applied
to standard non-relativistic quantum mechanics written in parametrized form.
These results also showed how spacetime questions in
non-relativistic quantum mechanics can be expressed in a fully spacetime form,
involving an inner product defined on spacetime.

We applied our formalism to some simple examples involving the free particle
in one and two dimensions. These examples showed that the class operators
could be easily calculated. Furthermore, there were some situations in which
decoherence was possible for special states, without the need for an environment.
However, the results were not always easy to interpret, and this is largely due
to the fact that the classical limit of the quantum theory brings in
reflecting potentials.

We briefly discussed the relativistic particle and considered the question of the
probability of crossing a spacelike surface. We found that only states antisymmetric
about the surface give decoherence and the crossing probability is then zero.
This means that our formalism does not appear to reproduce earlier heuristic
formulae for surface crossings. Furthermore, this result also emphasized that
our approach is in fact a genuinely different proposal for the class operators
compared to other approaches, and not a formalization of earlier more heuristic
ideas.

The formalism boiled down to particularly simple expressions for the case
of systems of non-interacting harmonic oscillators and we exhibited some simple
eigenstates which gave approximate decoherence and which had a clear semiclassical interpretation.

A future publication will revisit the calculations of this paper but with the inclusion
of an environment to produce decoherence \cite{HaWa}.

\section{Acknowledgements}

We are very grateful to Jim Hartle for many useful conversations.
This work was supported in part by EPSRC Grant EP/C517687/1.
P.W. was partially supported by the Leventis Foundation.

\bibliography{apssamp}

\begin{thebibliography}{10}
\bibitem{HaTh1} J.J.Halliwell and J.Thorwart, {Phys.Rev.} {\bf
D64}, 124018 (2001). [Note that Eq.(5.26) in this paper is
incorrect. The suggestion that modified
class operators may be obtained by modifying the sums
over the parametrization of the paths may still be true but
the proof given is not valid].
\end{thebibliography}

\end{document}